\documentclass{ws-procs9x6}

\begin{document}
\title{Spontaneous Scherk-Schwarz supersymmetry breaking and radion
stabilization
~\footnote{\uppercase{B}ased on plenary talks presented
at {\it \uppercase{SUSY} 2003: \uppercase{S}upersymmetry in the
\uppercase{D}esert}\/, held at the \uppercase{U}niversity of
\uppercase{A}rizona, \uppercase{T}ucson, \uppercase{AZ},
\uppercase{J}une 5-10, 2003; and at  {\it\uppercase{IPPP}
\uppercase{W}orkshop on:  \uppercase{S}tring
\uppercase{P}henomenology 2003}, held at the \uppercase{U}niversity of
\uppercase{D}urham, \uppercase{U.K.},\ \uppercase{J}uly
29-\uppercase{A}ugust 4, 2003.  \uppercase{T}o appear in the
\uppercase{P}roceedings.}
}

\author{MARIANO QUIROS}

\address{Theoretical Physics Group \\
ICREA-IFAE, \\ 
E-08193 Bellaterra (Barcelona) Spain\\ 
E-mail: quiros@ifae.es}


\maketitle

\abstracts{In this talk I review the issues of supersymmetry breaking
and radion stabilization in a five dimensional theory compactified on
the $\mathbb Z_2$ orbifold. Supersymmetry breaking by Scherk-Schwarz
boundary conditions is interpreted as spontaneous breaking of local
supersymmetry by the Hosotani mechanism. The auxiliary field
responsible for spontaneous supersymmetry breaking is inside the
five-dimensional off-shell minimal supergravity multiplet. Different
ways of fixing the supersymmetry breaking order parameter are
analyzed. In the presence of supersymmetry breaking the one-loop
effective potential for the radion has a minimum that fixes its vacuum
expectation value. The radion is stabilized in a metastable
Minkowski$_4$ minimum (versus the AdS$_4$ vacuum) with a mass in the
meV range making it interesting for future deviations from the
gravitational inverse-square law.}

\section{Introduction}

Higher dimensional theories share the general problem of how to fix
the radii of compact dimensions and, if they are supersymmetric, of
how to break supersymmetry. While the first question is dependent on
the second one, for in the absence of supersymmetry breaking the
radion potential is flat and its vacuum expectation value (VEV)
undetermined, the second question finds a very elegant solution in the
Scherk-Schwarz (SS) mechanism~\cite{ss} where supersymmetry is broken
by global effects~\cite{quiroslec}. In this talk I will review
possible solutions to both questions and what they imply for radion
phenomenology. I will use as a prototype a five-dimensional (5D)
theory compactified on a $\mathbb Z_2$ orbifold with a flat
geometry. This corresponds to 5D ``ungauged'' supergravity where the
gravitino does not have any tree-level mass-term either in the bulk or
localized at the boundaries.

\section{Scherk-Schwarz supersymmetry breaking as spontaneous breaking}

Scherk-Schwarz supersymmetry breaking of a 5D theory compactified on
$S^1/\mathbb Z_2$ can be interpreted as a spontaneous breaking of 5D
local supersymmetry~\cite{geromariano}. To interpret SS supersymmetry
breaking as a Hosotani mechanism~\cite{hosotani} one has to go to the
off-shell version of 5D N=1 supergravity where the $SU(2)_R$
automorphism of global supersymmetry is gauged by a tripet of
auxiliary fields $\vec{V}_M$. The ungauged version of this theory has
been recently formulated~\cite{Zucker:1999ej,Zucker:ww} and found that
two multiplets are necessary: the minimal supergravity multiplet
$(40_B+40_F)$ and a tensor multiplet $(8_B+8_F)$ with appropriate
parities dictated by the orbifold group invariance of the theory. The
only physical fields are in the gravitational multiplet: the graviton
$e_M^A$, the gravitino $\psi_M^i$ (where the index $i$ transform as an
$SU(2)_R$ doublet), and the graviphoton $B_M$. All other fields are
auxiliary: in particular the relevant fields for supersymmetry
breaking are the even fields $V_5^{1,2}$ that constitute the $F$-term
of the radion superfield. SS supersymmetry breaking has also been
studied in the context of gauged (AdS$_5$)
supergravity~\cite{Lalak:2002kx,Bagger:2003vc}.

In particular in the background of $V_5^{2}$ the Goldstino is provided
by the fifth component of the gravitino ($\psi_5$) as it is obvious
from the local supersymmetric transformation
%
$\delta_\xi \psi_5=\mathcal{D}_5\xi+\cdots=i\sigma^2 V_5^2\xi+\cdots$
%
A local supersymmetry transformation with parameter $\xi\equiv
-(\mathcal{D}_5)^{-1}\psi_5$ gauges $\psi_5$ away and gives a mass to
the gravitino. This defines the ``super-unitary'' gauge where $\psi_5$
has been ``eaten'' by the four-dimensional gravitino $\psi_\mu$. In
fact using the coupling of $V_5^2$ to the gravitino field through the
covariant derivative $\mathcal{D}_5$ one obtains gravitino mass
eigenvalues for the Kaluza-Klein modes where $m_{3/2}^{(0)}\propto
\langle V_5^2\rangle$ and $\langle V_5^2\rangle$ can be identified
with $\omega/R$ where $R$ is the physical radius of the extra
dimension and $\omega$ the SS parameter. I will next
review different procedures for fixing both the SS parameter $\omega$
and the physical orbifold radius $R$.

\section{Fixing the Scherk-Schwarz parameter}

In Ref.~\cite{vonGersdorff:2003rq} we described how to fix the SS
parameter using 5D off-shell supergravity tools. There we explained
how the tensor multiplet formalism and its dual, the linear multiplet
one, in 5D supergravity are not equivalent in the presence of a
non-trivial cohomology, as that possessing the orbifold $S^1/\mathbb
Z_2$. In particular using the tensor field $B_{MNP}$ (tensor
multiplet) and its field equation the one-form $V_M$ is closed
($dV=0$). Since the space is cohomologically non-trivial that form can
be non-exact. In that case the VEV $V_5^2$ has a non-vanishing, but
tree-level undetermined, Wilson flux as
%
$\frac{1}{2\pi}\oint dx^5 \langle V_5^2\rangle\equiv 2 \omega.$
%
In this case supersymmetry is broken but tree-level
undetermined. Supersymmetry can then be broken by radiative
corrections~\cite{geromarianotoni,rayner} as in the Hosotani breaking
of a gauge theory.  Of course this procedure does not violate any
non-renormalization theorem since the tree-level potential is flat and
for $\langle V_5^2\rangle=0$ the vacuum energy is zero.

In the linear multiplet formalism the tensor $B_{MNP}$ is traded by
the vector $W_M$. SS supersymmetry breaking is based on an intriguing
property of the linear multiplet. I will first concentrate on a vector
field $E_M$ with vanishing field strength, $dE=0$. This defines a
Maxwell multiplet where all other components are equal to zero. This
configuration is left invariant under local supersymmetry since local
supersymmetric transformations only depend on $E_M$ through its field
strength, $dE$. This multiplet has no physical degrees of freedom but
on non-simply connected spaces it can have a non-vanishing flux $\int
E_M dx^M$. We call it ``flux-multiplet''. Under $\mathbb Z_2$
compactification with even $E_5$ the flux multiplet reduces to the
constant multiplet that is known to be supersymmetric. It is possible
to fix the VEV of $V_5^2$ at tree level by using an independent source
of supersymmetry breaking that will play the role of the
superpotential in the low energy effective theory. This was done in
Refs.~\cite{Bagger:2001ep,Rattazzi:2003rj} by attaching this
superpotential to the branes,~i.e. by choosing the flux multiplet with
$E_M=\delta_M$, where
$\delta_5=\omega_0\delta(x^5)+\omega_\pi\delta(x^5-\pi R)$ and
$\delta_\mu=0$, that obviously satisfies the condition $dE=0$. We can
even generalize the source term by using any fixed closed form as the
constant
one~\cite{vonGersdorff:2003rq},~i.e. $\delta_5=\omega$. According to
the general analysis of Ref.~\cite{vonGersdorff:2003rq} there is
nothing special about the orbifold and we can use this particular
formalism to implement supersymmetry breaking on the circle. One
should of course worry that such term might be breaking general
coordinate invariance. In fact it does not since a fixed closed one
form is not the same in any frame but it differs by a non-physical
gauge transformation. Of course a different issue is the physical
origin of the flux multiplet, a point where we are not going to enter
here.

In both cases the radion potential is flat at tree-level and we need
the use of the Casimir energy to fix it. In
Ref.~\cite{vonGersdorff:2003rq} we have thoroughly analyzed the two
previous cases. In the case of the tensor multiplet formalism both the
SS parameter and the radion VEV have to be determined by the one-loop
effective potential. In that case the two-field minimization leads to
a VEV $\omega=1/2$, independently on the radion VEV. In the case of
the linear multiplet formalism $\omega$ is fixed at the tree-level by
whatever dynamics at the brane or bulk we like, and the one-loop
potential provides the radion VEV as a function of the SS
parameter. Since the tensor multiplet formalism is more involved and
the linear multiplet one allows for arbitrary values of $\omega$, in
the following we will concentrate on the latter case.

\section{Radion effective potential}

We will consider radion stabilization using the Casimir energy.  We
parametrize the 5D metric in the Einstein frame
as~\cite{Appelquist:1982zs}
%
$ds^2=G_{MN}dx^M dx^N\equiv\phi^{-\frac{1}{3}}g_{\mu\nu} dx^\mu
dx^\nu+\phi^{\frac{2}{3}} dy^2.$
%
where $y=x^5$ goes from $0$ to $L$. The radion field, whose VEV
determines the size of the extra dimension is $\phi^{\frac{1}{3}}$ and
the physical radius is given by $R=\langle\phi\rangle^\frac{1}{3} L$.
The length scale $L$ is unphysical and completely arbitrary. It will
drop out once the VEV of the radion is fixed and the effective 4D
theory will only depend on $R$.  In order to achieve zero
four-dimensional cosmological constant we will introduce bulk
cosmological constant $g^2$ and brane tensions $T_{0,\pi}$ as possible
counterterms.  This corresponds to AdS$_5$ supergravity, although the
AdS gauge coupling $g$ (as well as the brane tensions) are really one
loop counterterms and there is no tree level warping. The
four-dimensional effective Lagrangian including the radion one-loop
effective potential is $\mathcal L=-V+\pi L g^2 \phi^{-\frac{1}{3}}+
\frac{1}{2}(T_0+T_\pi)\phi^{-\frac{2}{3}}$, where $V$ is the Casimir
energy.

\subsection{Propagating bulk fields}

By considering $N_V$ vector multiplets and $N_h$
hypermultiplets propagating in the bulk, the Casimir energy
is~\cite{}
\begin{equation}
V_{eff}\propto(2+N_V-N_h)\frac{1}{L^4 \phi^2}.
\label{Veffnomass}
\end{equation}
Potential (\ref{Veffnomass}) is runaway and provides either a
repulsive or an attractive force. Of course by adding the counterterms
one can create a stable minimum. Unfortunately the required
counterterms are not consistent with supersymmetry. The way out is to
introduce a mass scale in the theory. This can be done either by
introducing a supersymmetric odd mass for some hypermultiplets (that
produces an exponential localization of their lightest eigenstate on
an orbifold fixed point) and/or by introducing localized kinetic
terms~\cite{Ponton:2001hq}.

\subsection{Localized effects}

The fields that are strictly localized on the boundary fixed points
are four-dimensional fields and as such they cannot influence the
Casimir energy. However bulk hyperscalars with brane mass terms and/or
localized kinetic terms can influence the bulk Casimir energy while
they introduce scales into the theory. In particular the bosonic
Lagrangian of a supersymmetric hypermultiplet $(\varphi^i,\psi)$ with a
localized odd parity mass term $M(y)=\eta(y)M$, where $\eta(y)$ is the
sign function on $S^1$ with period $\pi R$, and localized kinetic
terms can be written as~\cite{rayner}:
\begin{align}
\mathcal L=&\left|\mathcal D_M\varphi\right|^2-M^2(y)
\left|\varphi\right|^2+M^\prime(y)(\varphi^\dagger\sigma_3\varphi)\nonumber\\
+&\frac{2}{M}\left[c_0\delta(\pi)+c_\pi\delta(y-\pi
R)\right]\left(\partial_\mu\varphi\right)^2
\label{lag1}
\end{align}
where the coefficients $c_{0,\pi}$ have been normalized to be
dimensionless. The 4D mass spectrum is given by
\begin{align}
\sin^2\omega\pi=&\sin^2(\Omega\pi R)\
\frac{m^2(1-c_0+c_\pi)+c_0c_\pi m^4/M^2}{\Omega^2}\nonumber\\
+&\frac{m^2(c_0+c_\pi)}{2M\Omega
}\sin (2\Omega\pi R)
\label{masa2}
\end{align}
where $\Omega=\sqrt{m^2-M^2}$. 

The effective potential can be computed using standard
methods~\cite{delgado,geromarianotoni}. For the case of $N_H$
hypermultiplets with a common mass $M$ and $c_0=c_\pi=0$ it gives the
result
\begin{equation}
V_{eff}=M^6L^2\, \frac{N_H}{8}\int dz\, z^3
\ln\left[1+\frac{z^2+x^2}{z^2\sinh^2\left(\sqrt{z^2+x^2}\right)}\right]
\label{pot2}
\end{equation}
where $x=M \pi R$.

The total one-loop effective potential is a combination of potential
(\ref{Veffnomass}), corresponding to fields propagating in the bulk,
and potential (\ref{pot2}) corresponding to bulk
fields with localized effects.

\section{Radion stabilization}

For any value of the SS parameter $\omega$ the total potential for the
radion field has a global minimum that depends on $\omega$ and
$M$. If we now introduce bulk $g^2>0$ and tension $T_{0}+T_\pi>0$
counterterms fine-tuned to have zero cosmological constant and
consistent with 5D supersymmetry in AdS space, the radion minimum is
shifted to a value that depends on the counterterms. 
\begin{figure}[ht]
\centerline{\epsfxsize=3.7in\epsfbox{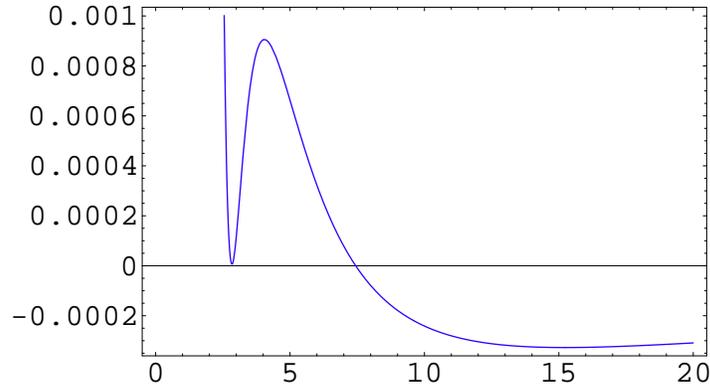}}
\caption{Radion potential for $\omega=0.25$ as a function of the
variable $M R \pi $, for appropriate values of the
counterterms. \label{inter}}
\end{figure}
On the other hand, as can be seen in Fig.~\ref{inter}, the counterterms
introduced to cancel the 4D cosmological constant produce an AdS$_4$
global minimum so that the stability of the Minkowski vacuum becomes a
real issue. 

The kinetic term for the field $\phi$ assumes the form $M_4^2
(\partial_\mu\phi)^2/\phi^2$ (where $M_4$ is the 4D Planck scale) and
the potential is $M^4f(\phi)$. In terms of the canonically normalized
radion $\varphi=M_4\ln\phi/\sqrt{3}$ the barrier separating the two
vacua, as well as the depth difference $\varepsilon$ between them, is
of order $M^4$ while the distance between both minima is of order
$M_4$. Despite the small barrier between the two vacua, in order to
tunnel a macroscopic bubble has to be nucleated and the energy cost
for it is huge, which leads to exponentially suppressed
probability. In fact we can model the potential around the metastable
minimum as $V(\varphi)= \lambda(\varphi^2-M_4^2)^2$, where
$\lambda\sim(M/M_4)^4$. The probability in the thin wall
approximation~\cite{Coleman:1980aw} is found to be $P\sim \exp[-B]$
where $B\sim M_4^{12}\lambda^2\varepsilon^{-3}\sim(M_4/M)^4\sim
10^{60}$. We can conclude that the Minkowski vacuum is stable on
cosmological times.

\section {Radion stabilization and the hierarchy problem}
Unlike in those approaches
where a warped geometry solves the hierarchy problem, in flat space we
must invoke supersymmetry for solving it. 
Our only concern was to obtain
a physical radius $< 1/$TeV. However this range is technically
natural since we are introducing bulk masses in the TeV range. A
different (not unrelated) issue is the origin of the weakness of
gravitational interactions in the 4D theory and its relation with
radion fixing. Here we have been working in a 5D gravity theory, with
a 1/TeV length radius, and therefore the presence of submillimeter
dimensions is not consistent with our mechanism for radion
stabilization. On the other hand the relation between the Planck
scales in the 4D and 5D theories, $M_4^2=M_5^3R$, with $R\sim 1/$TeV
implies that the scale where gravity becomes strong in the 5D theory
is much higher than $1/R$. This means that gauge interactions of the 5D
theory become non-perturbative at a scale $M_s\ll M_5$ in the
multi-TeV range. The theory should then have a cutoff at the scale
$M_s$ where a more fundamental theory should be valid. An example of
such behaviour is provided by Little String Theories (LST) at the
TeV~\cite{LST,Ant} where the string coupling $g_s\ll 1$ and $M_s$ and
$M_5$ are related by $M_5^3=M_s^3/g_s^2$. In other words $M_5$ does no
longer play the role of a fundamental field theoretical cutoff
scale. In these theories the weakness of the gravitational
interactions is provided by the smallness of the string
coupling. Moreover a class of LST has been found~\cite{Ant} where the
Yang-Mills coupling is not provided by the string coupling but by the
geometry of the compactified space where gauge interactions are
localized, e.g.~$g_{YM}\sim \ell_s/R$. Since the field theory has a
cutoff at $M_s$ the consistency of the whole picture relies on the
assumption that there is a wide enough range where the 5D field theory
description is valid.

\section{Radion phenomenology}

In the metastable vacuum the squared mass of the canonically
normalized radion field is given by $\sim$ (one-loop factor)$\times
M^4/M_4^2$. Since the size of the odd-mass term $M$ may be taken to be
of the order of 10 TeV, we conclude that the radion field acquires in
the metastable vacuum a mass around $(10^{-3}-10^{-2})$ eV. This range
of masses is interesting for present and future measurements of
deviations from the gravitational inverse-square law in the millimeter
range~\cite{Adelberger:2003zx}. In particular this shows that a
positive-signal in table-top gravitational experiments does not
necessarily implies the existence of sub-millimeter dimensions.

Finally, we should also be concerned about the backreaction of the
Casimir energy and the counterterms on the originally flat 5D
gravitational background.  A dimensional analysis shows that the
effect of the counterterms by themselves would result in a warp factor
with a functional dependence on the extra coordinate as $a(\epsilon
My)$, where $\epsilon=\mathcal O(M/M_5)^{3/2}\equiv \mathcal
O(M/M_4)\sim 10^{-15}$ for $M\sim$ TeV.  Such a warping is completely
negligible. One can also show that the size of the gravitino bulk and
brane masses generated by the counterterms are of the order of the
radion mass and thus negligible as compared to the size of
supersymmetry breaking contributions.

\section*{Acknowledgments}
I wish to thank my collaborators G.~v.~Gersdorff, L.~Pilo,
D.~A.~J.~Rayner, A.~Riotto, for so many discussions we had on this
field and the joint work.  This work was supported in part by the RTN
European Programs HPRN-CT-2000-00148 and HPRN-CT-2000-00152, and by
CICYT, Spain, under contracts FPA 2001-1806 and FPA 2002-00748.


\end{document}